\documentclass[]{aa}
\usepackage{natbib}
\usepackage{graphicx}
\usepackage{color}
\usepackage{ulem}

\begin{document}

\title{Heating of the solar photosphere during a white-light flare}

\author{Jan Jur\v{c}\'{a}k
        \inst{1}
        \and
        Jana Ka\v{s}parov\'{a}
        \inst{1}
        \and
        Michal \v{S}vanda
        \inst{1,2}
        \and
        Lucia Kleint
        \inst{3, 4}}

\institute{Astronomical Institute of the Czech Academy of Sciences, Fri\v{c}ova  298, 25165 Ond\v{r}ejov, Czech Republic
\and 
  Astronomical Institute, Charles University, V~Hole\v{s}ovi\v{c}k\'ach 2, 18000 Praha, Czech Republic
  \and
  University of Applied Sciences and Arts Northwestern Switzerland, Bahnhofstrasse 6, 5210 Windisch, Switzerland
  \and
  Kiepenheuer-Institut f\"{u}r Sonnenphysik, Sch\"{o}neckstr. 6, 79104 Freiburg, Germany
  }

\date{Received 11 December, 2014; accepted }

\abstract
{The \ion{Fe}{i} lines observed by the Hinode/SOT spectropolarimeter were always seen in absorption, apart from the extreme solar limb. Here we analyse a unique dataset capturing these lines in emission during a solar white-light flare.}
  {We analyse the temperature stratification in the solar photosphere during a white-light flare and compare it with the post-white-light flare state.}
 {We used two scans of the Hinode/SOT spectropolarimeter to infer, by means of the LTE inversion code Stokes Inversion based on Response function (SIR), the physical properties in the solar photosphere during and after a white-light flare. The resulting model atmospheres are compared and the changes are related to the white-light flare.}
{We show that the analysed  white-light flare continuum brightening is probably not caused by the temperature increase at the formation height of the photospheric continuum. However, the photosphere is heated by the flare approximately down to $\log \tau = -0.5$ and this  results in emission profiles of the observed \ion{Fe}{i} lines. From the comparison with the post-white-light flare state of the atmosphere, we estimate that the major contribution to the increase in the continuum intensity  originates in the heated chromosphere.}
{}
  

\maketitle

%
%

\section{Introduction}

Solar flares are widely believed to be a consequence of reconnection of the coronal magnetic field in a peculiar configuration. The magnetic energy stored in the entangled coronal loops is released suddenly during the flare and a large portion of the flare energy is radiated away in a wide range of wavelengths emerging from the intensively heated flare atmosphere. 

In the visible range of wavelengths the usual line emission is often accompanied by enhancement of continuum radiation, and such flares are called white-light flares \citep[WLFs, see e.g.][]{1966SSRv....5..388S,1989SoPh..121..261N}. There are various mechanisms proposed for enhancement of the optical continuum: hydrogen bound-free and free-free transitions, Thomson scattering, and H$^{-}$ emission. Futhermore, each mechanism may dominate in different atmospheric layers spanning from the photosphere through the temperature minimum region to the chromosphere, and all require an increase in temperature and electron density in those layers.

However, it is still debated how these layers are heated. Several processes have been proposed:  electron and/or proton bombardment, XEUV heating, Alfv\'{e}n wave dissipation, etc. \citep[see e.g.][]{1990ApJ...365..391M}. Moreover, it has been shown that the photosphere and the chromosphere can be radiatively coupled via photospheric heating by  H$^{-}$ absorption of the hydrogen Balmer continuum, which originates in the chromosphere. This backwarming then can lead to increased photospheric (H$^{-}$) radiation \citep[e.g.][]{1989SoPh..124..303M}.

To disentangle the contributions to the visible-light continuum, specific observations  and dedicated models are needed. Combining flare observations of several photospheric and chromospheric lines together with visible-light continuum and non-LTE modelling, \citet{1990ApJ...360..715M} constructed a semi-empirical model of a WLF. Recently, using off-limb flare observations in the HMI/SDO pseudo-continuum and non-LTE RHD approach, \citet{Heinzel:2017} have reported the presence of the hydrogen Paschen continuum originating in the chromosphere. Other HMI off-limb sources related to a flare were found to be of two kinds (chromospheric and coronal), and interpreted as free-bound continuum (and possible line emission) and Thomson scattering, respectively \citep{Martinez:2014,Saint-Hilaire:2014}. Furthermore, the hydrogen Balmer continuum was observed during flares by IRIS \citep{2014ApJ...794L..23H, 2016ApJ...816...88K,2017ApJ...836...12K}. Additionally, near-infrared emission at $1.56 \mu m$ was detected during several X-class WLFs \citep{2006ApJ...641.1210X,2012ApJ...750L...7X}. In  an undisturbed solar atmosphere this emission is considered to originate at the opacity mininum located  below $\tau_{500}=1$ (where $\tau_{500}$ is the optical depth at 500~nm).

The Solar Optical Telescope  \citep[SOT,][]{Tsuneta:2008} aboard the Hinode satellite \citep{Kosugi:2007} has provided observations in broad-band filters as well as spectropolarimetry of a pair of photospheric \ion{Fe}{i} lines. Continuum broad-band filters have been used to detect and analyse WLF emission \citep[e.g.][]{2014ApJ...783...98K}. Here, we focus on spectropolarimetric data. The spectropolarimeter attached to the SOT has been measuring the Stokes profiles of the \ion{Fe}{i} 6301.5 and 6302.5 \AA\ lines since 2006. Analyses of these data have given us great insight into the structure of the magnetic field in the solar photosphere, both in quiet-Sun and active regions. In all cases, these \ion{Fe}{i} lines were observed in absorption on the solar disk. This is caused by a temperature decrease with height, and thus a decrease in  the source function, in the layers of the solar photosphere where these lines are formed. Only in the case of observations at the extreme solar limb, were these lines  detected in emission \citep{Lites:2010}.{\color{red}}

After 11~years of continuous observations, we were lucky enough that the raster scan of the Hinode spectropolarimeter crossed a WLF ribbon and observed there emission profiles that allow us to study in detail the response of the solar photosphere to the X-class flare. These unique observations were already used to test the reliability of the HMI pseudo-continuum intensities and to assess the suitability of that HMI product for photometry during WLFs \citep{Svanda:2018}.

In this paper we determine the photospheric temperature structure during a WLF observed on September 6, 2017, by means of inversion of full Stokes profiles and estimate the photospheric and  chromopheric contributions to the observed visible continuum.

\begin{figure*}[!t]
 \centering \includegraphics[width=0.85\linewidth]{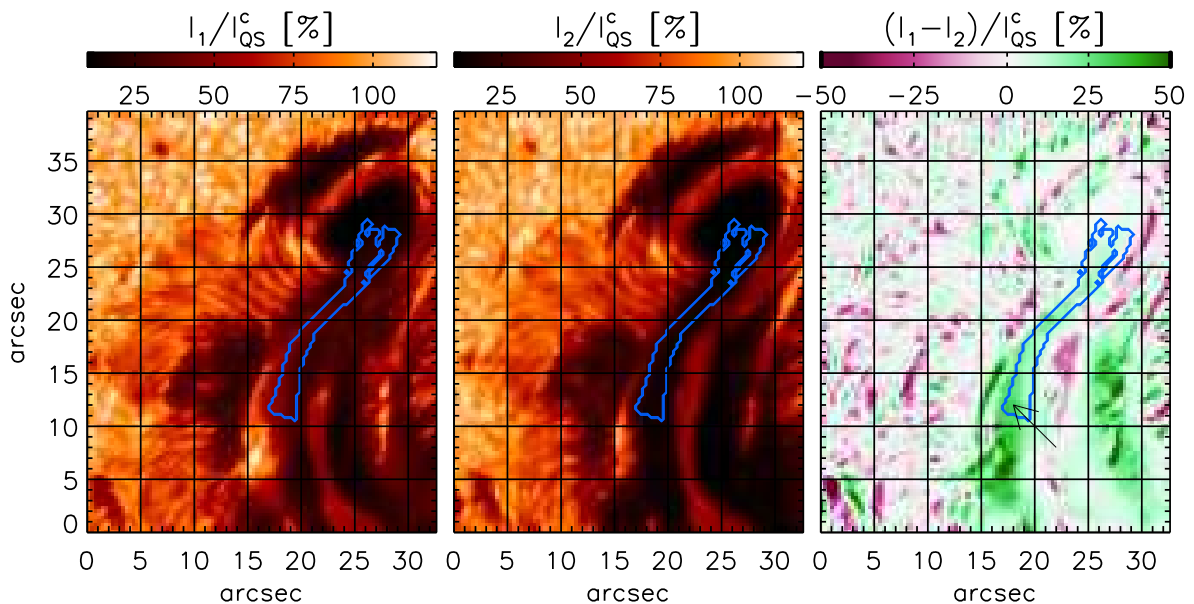}
 \caption{Continuum intensity maps reconstructed from the two Hinode raster scans. The left map was scanned between 11:57~UT and 12:04 UT and captured a WLF ribbon; the middle map was scanned between 12:19~UT and 12:42~UT. On the right is the intensity difference between these two scans. The blue contour indicates the region where we ascribe the intensity difference to the WLF. The arrow points to a pixel where we observed the Stokes profiles displayed in Figs.~\ref{profile_comparison} and~\ref{profile_reduction}. The labels  $I_1$ and $I_2$  correspond to the continuum intensities observed during the first and second Hinode/SP scan, respectively.}
 \label{hinode_int}
\end{figure*}

\begin{figure*}[!t]
 \centering \includegraphics[width=0.85\linewidth]{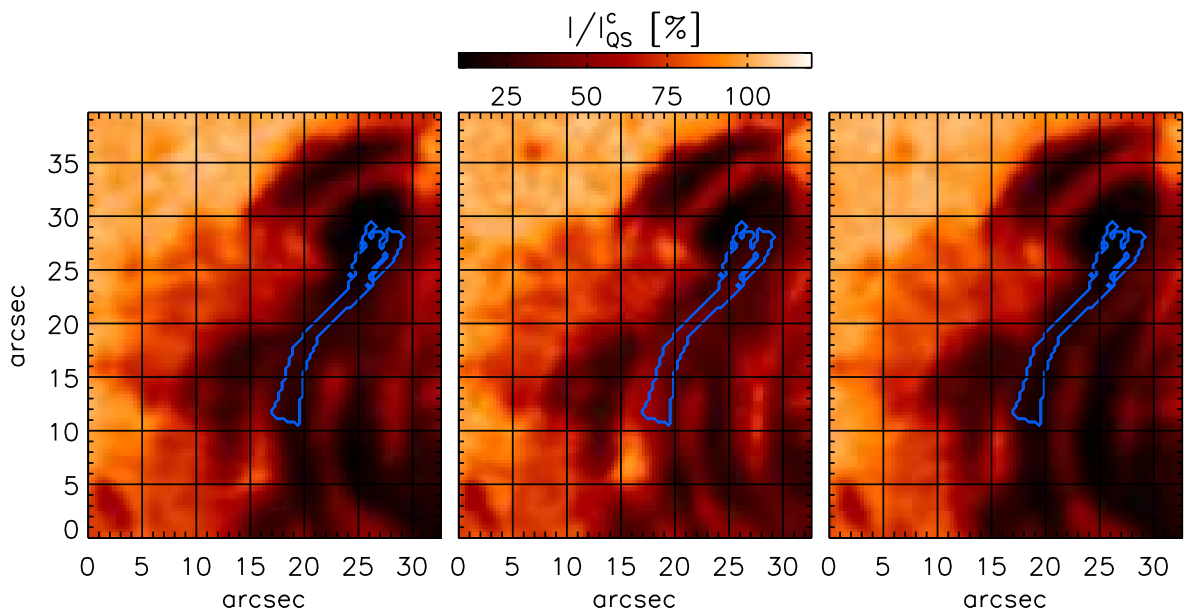}
 \caption{Pseudo-continuum intensity maps from SDO@HMI. We show the pre-flare phase at 11:31~UT (left), flare phase at 12:01~UT (middle), and post-WLF phase at 12:31~UT (right). For display purposes, the HMI data were interpolated to the Hinode SP resolution.}
 \label{HMI_int}
\end{figure*}

\section{Observations and data analysis}
\label{observations}

\begin{figure*}[!t]
 \centering \includegraphics[width=0.8\linewidth]{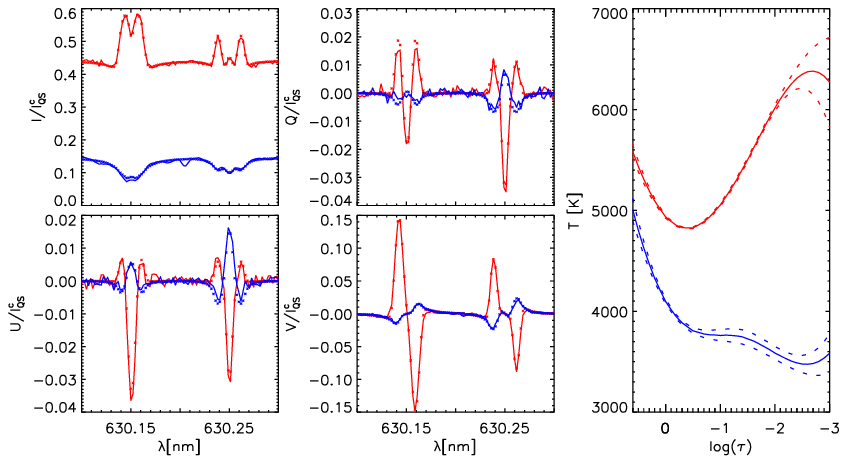}
 \caption{Left, small plots: Comparison of the Stokes profiles observed during the WLF (red lines) and in the post-WLF phase (blue lines); the $*$ symbols in the respective colours indicate the best fit of these profiles achieved with the inversion code. Right: Temperature stratifications obtained by the inversion code for the flare (red) and post-WLF (blue) phase; the dashed lines show the error margin determined by the inversion code SIR.} 
 \label{profile_comparison}
\end{figure*}

We use two raster scans of the active region NOAA~12673 taken by the spectropolarimeter (SP) attached to Hinode@SOT. At the time of the observations the active region was located 34$^\circ$ west and 14$^\circ$ south from the solar disk centre. 

The first scan captured part of the active region with the WLF ribbon of the X9.3-class flare that occurred on September 6, 2017 (SOL2017-09-06T11:53). The second scan was taken approximately 30~min later and we use it as a reference for a solar photosphere not influenced by a flare. 

Hinode SP records full Stokes profiles of the neutral iron line pair at 630 nm. The observed line profiles were calibrated using the standard reduction routines \citep{Lites:2013}. The first SP scan was taken in the  fast mode, for which the spatial sampling is 0\farcs32 along the slit and 0\farcs3 in the scanning direction. The second scan was taken in the normal mode, where the spatial sampling is two times better, i.e. 0\farcs15. The noise level  in both types of SP scans is around $2 \times 10^{-3}~I^c_{QS}$ (continuum intensity of the surrounding quiet Sun) for all Stokes profiles.

The second scan was rebinned by $2 \times 2$ pixels to the spatial sampling of the first scan and aligned with it using the method described in \citet{Luhe:1983}. This alignment method achieves sub-pixel precision, but there are local misalignments due to the sunspot evolution (see Sect.~\ref{results}).

To determine the physical properties of the solar photosphere, we used the code Stokes Inversion based on Response function \citep[SIR;][]{Cobo:1992}. Except for the temperature, all atmospheric parameters were considered height independent. We took into account the spectral point spread function of the Hinode SP in the inversion process. We assumed the magnetic filling factor to be unity and assumed no stray light. The macroturbulence was set to zero, while microturbulence was a free parameter of the inversion. To account for the complex emission profiles observed in the flare ribbon, we allowed  the temperature to change at five optical depths (nodes at $\log \tau_{500} =$ 1, -0.2, -1.4, -2.6, and -3.8; hereafter we drop index 500). The temperature stratification on a finer grid of optical depths was interpolated using splines between these nodes. The retrieved magnetic field azimuth ($\psi$) and inclination ($\gamma$) are in the line-of-sight (LOS) frame. We removed the $180^\circ$ azimuth ambiguity using the AMBIG code \citep{Leka:2009}. The LOS velocity ($v_\mathrm{LOS}$) is calibrated such that the average $v_\mathrm{LOS}$ in the sunspot umbrae is zero.

We note that the Stokes profiles having an area asymmetry cannot be fitted by the assumed atmospheric structure. With the spatial resolution of Hinode SP, these profiles are commonly observed in certain regions of sunspot penumbrae and light bridges when observed off disk centre. However, the Stokes profiles in the umbral region that we focus on are symmetric and the simple model of atmosphere is justifiable. Also, the SIR code works under the assumption of LTE and hydrostatic equilibrium. We are certain that there are effects that cannot be correctly accounted for by the SIR code especially during the flare phase; we discuss the reliability of the results in Sect.~\ref{discussion}.

\section{Results}
\label{results}

In Fig.~\ref{hinode_int}, we show the continuum intensity maps reconstructed from the two analysed raster scans (left and middle panels) and their difference (right panel). The peak values of the intensity differences may be ascribed to the misalignment of the two scans and to the evolution of the fine structures at the edges of light bridges, penumbrae, and pores. These areas are therefore not used for the detailed analysis, even though some of their enhancements are due to the flare because the lines appear in emission. 

There are also regions, where the intensity difference is caused by the WLF ribbon. The largest of these regions is indicated in Fig.~\ref{hinode_int} by a blue contour and our analysis is focused  on it. This region was selected manually, and pixels where the observed Stokes profiles were not reproduced successfully in the two SP scans (based on the $\chi^2$ value of the fit) were removed from this region. A second flare ribbon can be identified between [31\arcsec, 2\arcsec] and [27\arcsec, 19\arcsec], but we do not discuss it here because its major part is significantly influenced by the evolution of sunspot fine structure. 

To confirm that the  intensity change in this region is not a result of the morphological evolution of the sunspot light bridges, we also checked the HMI pseudo-continuum images including the pre-flare phase when Hinode data are not available. These intensity images are shown in Fig.~\ref{HMI_int}. The times of the HMI observations of the flare and post-WLF phase correspond roughly to times when the region indicated by the blue contour in Fig.~\ref{hinode_int} was scanned. It is clear that this region is bright only during the flare, and dark before and afterwards. \citet{Svanda:2018} showed that HMI pseudo-continuum maps can be successfully used for the detection of the WLF region, although they are not reliable regarding the quantitative changes in the continuum intensity.

In Fig.~\ref{profile_comparison}, we show an example of Stokes profiles observed in the region of the WLF ribbon (red) and compare them with the post-WLF Stokes profiles (blue) at the same spatial location [18\arcsec, 12\arcsec]. This pixel is located in the area with the maximum continuum intensity difference between the WLF and post-WLF phase in the studied region. In both cases, the inversion code SIR successfully fitted the observed Stokes profiles.

The major difference between the model atmospheres fitting the Stokes profiles in the flare and the post-WLF phase is the temperature stratification shown in the right panel of Fig.~\ref{profile_comparison}. During the flare phase, the atmosphere appears to be hotter in all layers compared to the post-WLF phase (the credibility of this result is discussed in Sect.~\ref{discussion}). 

The emission profiles can be explained by an increase in the temperature at higher layers of the line-forming region. In the particular case shown in Fig.~\ref{profile_comparison}, the temperature is increasing from $\log \tau = -0.5$ to the top of the line-forming region. The temperature decrease around $\log \tau = -2.8$ cannot be trusted as the inverted lines are no longer sensitive to the physical properties at these atmospheric heights. 

\begin{figure*}[!t]
 \centering \includegraphics[width=0.85\linewidth]{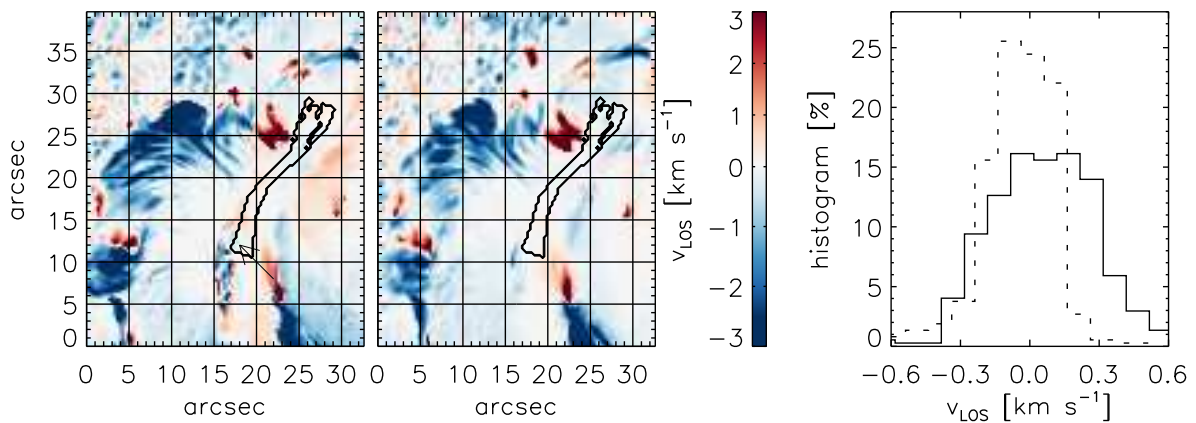}
 \caption{LOS~velocity maps for the WLF phase (left map) and the post-WLF phase (right map). Positive values (redshifts) correspond to downflows. The histograms on the right show the distribution of $v_\mathrm{LOS}$ in the analysed region where the solid and dashed lines correspond to the WLF and post-WLF phases, respectively. The arrow points to the pixel, where we observed the Stokes profiles displayed in Figs.~\ref{profile_comparison} and~\ref{profile_reduction}.}
 \label{velocity_maps}
\end{figure*}

The increased temperature in the higher photosphere also explains the stronger $Q$, $U$, and $V$ signals in the flare phase compared to the post-WLF phase as the amplitudes of the polarisation profiles are highly sensitive to the temperature. The opposite gradient of the temperature (and thus also source function under the assumed LTE) in the line-forming region, compared to the commonly found decrease, also causes the reversed shape of the Stokes $Q$, $U$, and $V$ profiles. To explain this shape reversal by a change in magnetic field orientation, we would have to suppose an unlikely scenario that,  between the flare and post-WLF scans, the inclination and azimuth changed by $180^\circ$ and $90^\circ$, respectively, with respect to the LOS.

The increasing temperature with height can also explain why we find stronger polarisation signals in the \ion{Fe}{i} 630.15~nm line than in the 630.25~nm line, which is more sensitive to the magnetic field. We note that in sunspots, the vast majority of the $Q$, $U$, and $V$ amplitudes of the 630.25~nm line are stronger than those of the 630.15~nm line. The amplitudes of the polarisation signals are given not only by the magnetic field strength, but also by temperature and since the \ion{Fe}{i} 630.15~nm line  forms slightly higher in the atmosphere and is more sensitive to the temperature than the \ion{Fe}{i} 630.25~nm line (see Fig.~\ref{rf}), its $Q$, $U$, and $V$ amplitudes may be stronger due to the higher temperatures at higher layers. An alternative explanation for this atypical ratio of polarisation signals in the \ion{Fe}{i} 630.15~nm and 630.25~nm lines (under LTE conditions) would be a steep gradient of magnetic field strength with height that is not likely in umbrae. Such a peculiar configuration of magnetic field was found only in the sunspot light bridges \citep{Jurcak:2006}. 

\begin{table}[!b]
\caption{Comparison of atmospheric model parameters in the flare and post-WLF phases for a pixel displayed in Fig.~\ref{profile_comparison}}
\label{model_comparison}
\centering 
\begin{tabular}{l c c c c}

phase & $B$ [G] & $v_\mathrm{LOS}$ [m s$^{-1}$] & $\gamma$ [deg] & $\psi$ [deg] \\
\hline 
post-WLF & $2490 \pm 210$ &  $60 \pm 300$ & $131 \pm 6$ & $115 \pm 7$ \\
flare & $2510 \pm 50$ & $40 \pm 90$ & $149 \pm 2$ & $105 \pm 3$ \\
\hline
\end{tabular}
\end{table}

In Table~\ref{model_comparison} we show that all other relevant physical parameters are comparable for the flare and post-WLF phases. There is a change in magnetic field inclination of 18$^\circ$ that cannot be ascribed to the uncertainties of $\gamma$. The field became more horizontal in this pixel and this result is in agreement with studies of  \citet{Gosain:2012} and \citet{Petrie:2012,Petrie:2016}, among others,  who investigated the magnetic field reconfiguration in the photosphere associated with flares. We note that larger statistical studies show that the field is equally likely to increase or decrease \citep[e.g.][]{2018ApJ...852...25C}. Similarly, we find a change in the magnetic field azimuth of 10$^\circ$, but such a change is not distinct enough and can be caused by the uncertainties of the retrieved $\psi$ values. 

We want to point out that there is no significant impact of the WLF on the LOS~velocity observed in the solar photosphere. This is not only the case of the $v_\mathrm{LOS}$ inferred in the pixel depicted in Table~\ref{model_comparison}. In Fig.~\ref{velocity_maps}, we show the maps of LOS~velocities derived from the two SP scans. The histograms on the right show that during the WLF phase there is an excess of redshifted pixels compared to the post-WLF phase. The amplitude of these downflows is typically below 0.5~km~s$^{-1}$.

\begin{figure}[!t]
 \centering \includegraphics[width=1\linewidth]{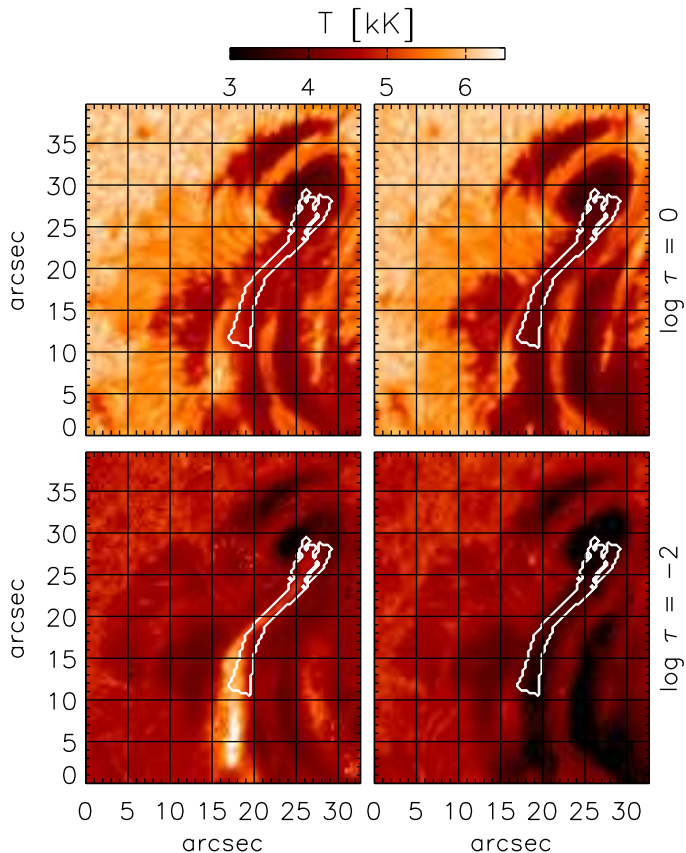}
 \caption{Temperature maps at optical depths $\log \tau = 0$ (top row) and $-2$ (bottom row) for the first (left column) and second (right column) Hinode/SP raster scan. The reliability of these temperature maps is discussed in Sect.~\ref{discussion}.} 
 \label{temperature_maps}
\end{figure}

\begin{figure*}[!t]
 \sidecaption
 \includegraphics[width=0.7\linewidth]{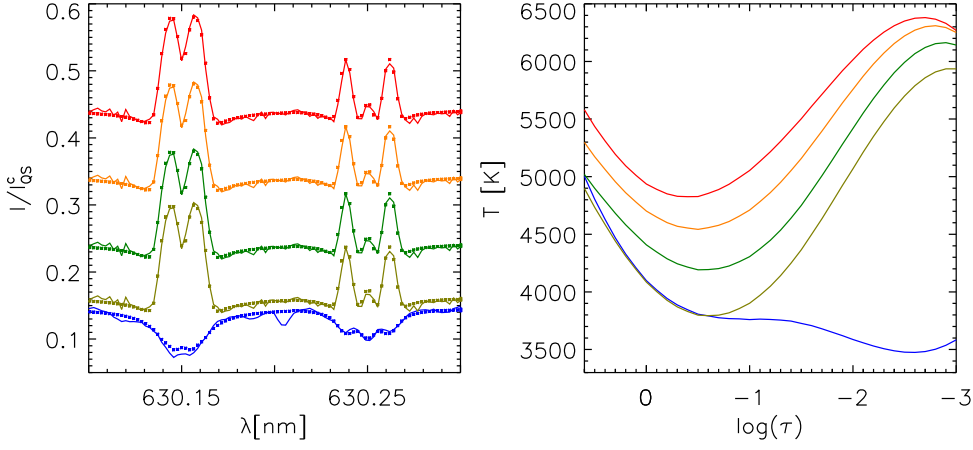}
 \caption{Comparison of the temperature stratifications resulting from the inversion of Stokes profiles, where the Stokes $I$ profile was reduced by a flat continuum of $0 I^c_{QS}$ (red), $0.1 I^c_{QS}$ (orange), $0.2 I^c_{QS}$ (green), and $0.28 I^c_{QS}$ (olive). The blue Stokes $I$ profile corresponds to the post-WLF phase at the same location and the temperature stratification indicated by the blue line corresponds to this profile. Left: Lines correspond to the observed profiles, and the $*$~symbols  to the best fits obtained by the inversion code SIR.} 
 \label{profile_reduction}
\end{figure*}
\begin{figure*}[!ht]
 \sidecaption
 \includegraphics[width=0.7\linewidth]{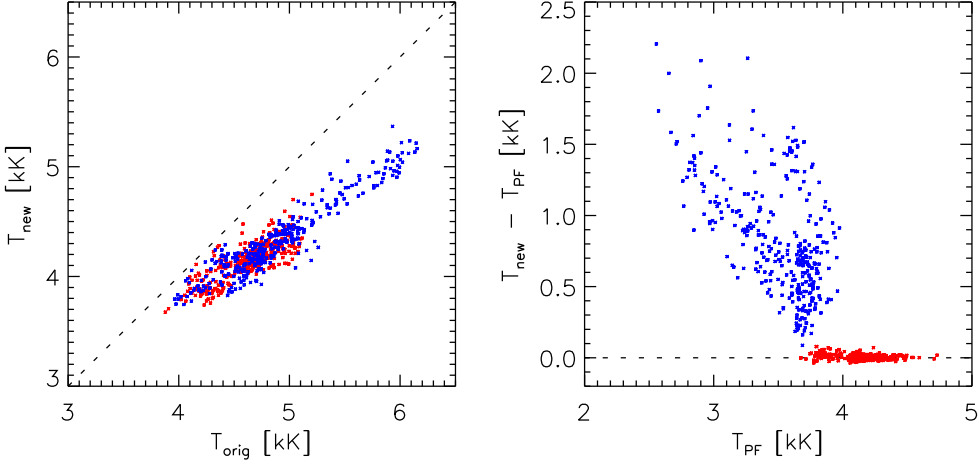}
 \caption{Left panel: Scatter plot showing the relation between the temperature resulting from the original inversion ($T_\textrm{orig}$) and temperature resulting from the Stokes $I$ profile reduced by a flat continuum ($T_\textrm{new}$). Right panel: Scatter plot showing the temperature change between $T_\textrm{new}$ and the temperature obtained from the post-WLF phase $T_\textrm{PF}$. Red and blue symbols correspond to $\log \tau = 0$ and $-2$, respectively.} 
 \label{scatterplot}
\end{figure*}

Temperature stratifications analogous to those shown in Fig.~\ref{profile_comparison} are obtained in the whole field of view pixel by pixel. Thus, they allow us to construct the temperature maps at optical depths of $\log \tau = 0$ and $-2$ that are shown in Fig.~\ref{temperature_maps}. The enhancement of temperature in the WLF ribbon is more obvious at higher photospheric layers, where it is created by a temperature increase analogous to the one shown in Fig.~\ref{profile_comparison}. 

Figure~\ref{temperature_maps} clearly indicates the effect of the WLF on the temperature at higher photospheric layers under the assumption of LTE. This increase in temperature is a feasible mechanism that can explain the Stokes profiles observed in the region of the WLF ribbon (Fig.~\ref{profile_comparison}). However, there are reasons to question the absolute values of the temperatures retrieved by the inversion code SIR. 

\section{Discussion}
\label{discussion}
\subsection{Evidence for the chromospheric-continuum enhancement}

The blue contour in Fig.~\ref{hinode_int} indicates a region where the intensity enhancement is caused by the WLF ribbon and not by the evolution of the sunspot fine structure. As shown in Fig.~\ref{profile_comparison}, the inversion code explains this intensity difference by enhanced temperatures at all optical depths, i.e. predicts also the heating of the solar photosphere around $\log \tau =0$, where the photospheric continuum forms. For the pixel shown in Fig.~\ref{profile_comparison}, the continuum enhancement of $0.3 I^c_{QS}$ is achieved by an increase in temperature at $\log \tau =0$ by 840~K compared to the post-WLF phase. 

Such an increase in temperature at the deepest photospheric layers is unlikely for several reasons. First, the minimum of temperature stratification in any of the investigated pixels is not below $\log \tau =-0.5$. Second, the observed emission profiles are never in pure emission. There is always a slight decrease in intensity in the far wings of the \ion{Fe}{i} 630.15~nm and 630.15~nm lines that indicates the temperature decrease above $\log \tau = 0$. Furthermore, if the temperature increase below the solar surface is real, the heating will have to be of a specific type, e.g. increase with depth because the atmospheric density significantly rises at those layers. 

Our assumption that the flare atmosphere is probably not altered significantly around $\log \tau = 0$ is further supported by the semiempirical flare models F1, F2, F3, F1$^{*}$ of \citet{1980ApJ...242..336M,1989SoPh..124..303M} and the FLA and FLB models for WLF \citep{1990ApJ...360..715M}. These coincide with the quiet-Sun  model C of \citet{1981ApJS...45..635V}  at heights $z < 0$~km, i.e. at $\log\tau>0$ \citep[see Fig.~1 in][]{1989SoPh..124..303M}. Also, \citet{2010ApJ...711..185C} modelled continuum emission in a sunspot atmosphere heated by non-thermal electron beams. Their results indicate that the temperature below $z=0$~km increases not more than by 100~K for one of the strongest beam heating they used.

\begin{figure*}[!t]
 \centering \includegraphics[width=0.85\linewidth]{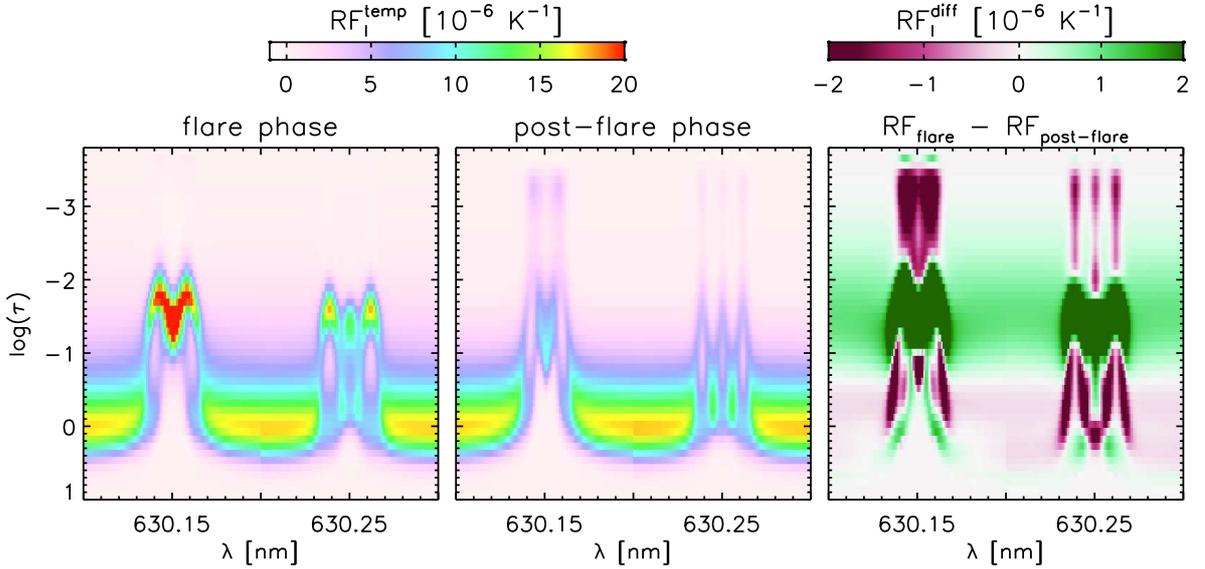}
 \caption{Response functions of the Stokes $I$ profile to the temperature for the model atmosphere fitting the emission profile shown in Fig.~\ref{profile_reduction} (left panel, in olive green) and for the model atmosphere fitting the post-WLF profile  in Fig.~\ref{profile_reduction} (middle panel, in blue). Their difference is shown in the right panel.} 
 \label{rf}
\end{figure*}

\begin{figure*}[!t]
\sidecaption
 \centering \includegraphics[width=0.7\linewidth]{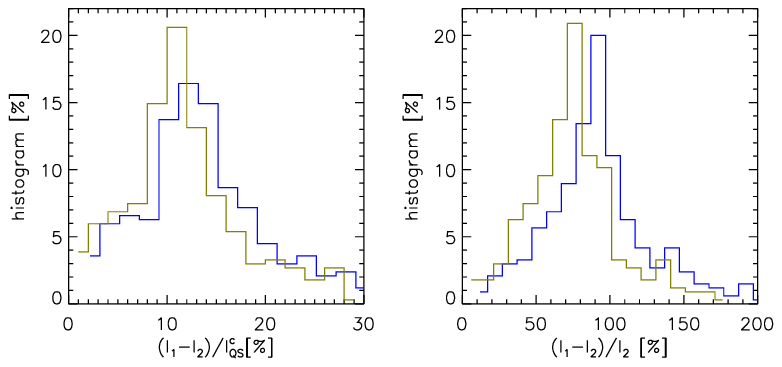}
 \caption{The histograms of the measured intensity increase between the flare and post-WLF states (blue, equivalent to the difference of continuum intensity of the red and blue profiles in Fig.~\ref{profile_reduction}) and the determined chromospheric continuum contribution to this increase (olive, equivalent to the difference of continuum intensity of the red and olive profiles in Fig.~\ref{profile_reduction}), both  assessed in the region of interest shown by the blue contour in Fig.~\ref{hinode_int}. Left: the values normalised by the continuum intensity of the quiet Sun $I^c_{QS}$, right: the values normalised by the local continuum intensity in the post-WLF state $I_2$.} 
 \label{histo}
\end{figure*}

Moreover, \citet{Heinzel:2017} explain observations of off-limb visible continuum sources during a M7.7 and a M1.7 flare. The authors conclude that the dominant source of the continuum radiation is  the Paschen recombination continuum and there are also smaller contributions by  Thomson scattering and by the hydrogen free-free emission. The continuum radiation from these sources is located in the solar chromosphere, i.e. well above the line-forming region of the \ion{Fe}{i} 630.15~nm and 630.15~nm lines. In the studied cases of M-class flares, the off-limb continuum intensity was around $0.1 I^c_{QS}$. 

Naturally, the inversion code SIR cannot account for such potential sources of  chromospheric continuum emission as it is applied to a photospheric line. Instead, SIR compensates for the continuum rise by increasing the temperature around the photospheric continuum-formation layer. To investigate whether such a contribution of continuum intensity from the solar chromosphere is realistic, we performed a set of inversions on the Stokes profiles observed in the region encircled by the blue contour in Fig.~\ref{hinode_int}. In order to mimic such a chromospheric optically thin contribution, we artificially decreased the Stokes $I$ intensity by a flat continuum, where for each pixel the observed WLF continuum was decreased  to the continuum intensity observed in the post-WLF phase (the step of the decrease was $0.01 I^c_{QS}$). This fine step of decreasing the $I^c_{QS}$ was used to  find the best match of temperature stratifications for the flare and post-WLF phases at each pixel, and  does not necessarily mean a matching continuum intensity. We suppose that the continuum contribution from the chromosphere is unpolarised, i.e. the Paschen continuum; therefore, the Stokes $Q$, $U$, and $V$ profiles remained the same.

In Fig.~\ref{profile_reduction} we show the results of these inversions for the pixel shown in Fig.~\ref{profile_comparison}, where the red and blue lines are identical in these plots. It is clear that the inversion code SIR can also reliably fit  Stokes profiles when we artificially decrease the Stokes $I$ profile by a flat continuum. Examples of such decreased Stokes $I$ profiles are shown in the left plot in Fig.~\ref{profile_reduction}; the  corresponding temperature stratifications are shown in the right plot. The other physical parameters of the model atmosphere are not affected significantly by the modification of the Stokes $I$ profile. 

We searched for the best match between model temperatures obtained from the inversions of the reduced Stokes $I$ profiles and the post-WLF temperature stratification in the same pixel in the interval of $\log\tau=[0.3,-0.3]$. For the case of the pixel shown in Fig.~\ref{profile_reduction}, the best match is displayed by olive lines, and the temperature stratifications  match within their uncertainties  in the range of optical depths from $\log \tau =0.7$ to $-0.7$. This is the range of optical depths where most of the photospheric continuum  forms.
 
The same approach can be used for all pixels encircled by the blue contour in Fig.~\ref{hinode_int}. Thus, we can estimate the bias of the temperature values in the flare ribbon displayed in Fig.~\ref{temperature_maps}. In the left plot of Fig.~\ref{scatterplot} we show how much the temperature changes between the original inversions and those resulting from the reduced Stokes $I$ profiles ($T_\mathrm{new}$). In all pixels and in both optical depths the original temperatures are overestimated. For the darkest pixels in the studied region the temperature difference is around 200~K, but this overestimation increases linearly with the temperature;  its maximum is around 500~K at $\log \tau =0$ and 1000~K at $\log \tau = -2$.

In the right plot of Fig.~\ref{scatterplot}, we show the absolute values of the temperature increase for the pixels in the studied region by comparing the temperatures resulting from the inversions of the post-WLF scan ($T_\mathrm{PF}$) with $T_\mathrm{new}$. This temperature difference has to be around zero at $\log \tau =0$ (red points) as we optimised the Stokes $I$ reduction to match $T_\mathrm{PF}$ and $T_\mathrm{new}$. At $\log \tau = -2$, the most common increase in temperature is between 300~K and 1000~K (the peak of the distribution is at 600~K). The maximum of the temperature increase is around 2000~K, but this occurs in very few pixels and is probably not reliable. For the pixel displayed in Fig.~\ref{profile_reduction}, the upper photosphere is 1500~K hotter compared to the post-WLF phase and this is, in the studied dataset, the highest reliable value of temperature increase in the upper photosphere. 

As implied before, the reduction of the Stokes $I$ profile by a flat continuum is optimised to achieve the match of the temperature value at $\log \tau =0$ for the flare and post-WLF phases. This does not necessarily mean that the continuum intensity would be the same for the reduced Stokes $I$ profile and the profile observed during the post-WLF phase. In the case of the pixel shown in Fig.~\ref{profile_reduction}, we find a discrepancy of $0.02 I^c_{QS}$ between the continuum intensity of the olive-coloured emission profile and the blue post-WLF profile. This is caused by a small contribution to the continuum intensity from the upper layer of the solar photosphere that is heated during the flare phase. 

This is clearly seen in Fig.~\ref{rf}, where we show the response functions of the two compared Stokes $I$ profiles to the changes in their respective temperature stratifications. The difference between these response functions (shown in the right panel of Fig.~\ref{rf}) indicates that  in the case of the emission profile, the photospheric continuum also forms  in higher photospheric layers, mostly between $\log \tau =-1$ and $-2$. 

The histograms of continuum intensity increases normalised to the intensity of the quiet Sun (left) and the local post-WLF intensity (right) are displayed in Fig.~\ref{histo} (blue curves). We note that the increase with respect to the local post-WLF continuum is typically around 100\%. We compare these observed increases with the derived chromospheric continuum enhancements (olive histograms in Fig.~\ref{histo}). The histograms of the chromospheric continuum contributions are shifted towards lower values, i.e. there is more than the flat chromospheric continuum to explain the observed continuum intensity increase.

\begin{figure}[!t]
 \centering \includegraphics[width=1\linewidth]{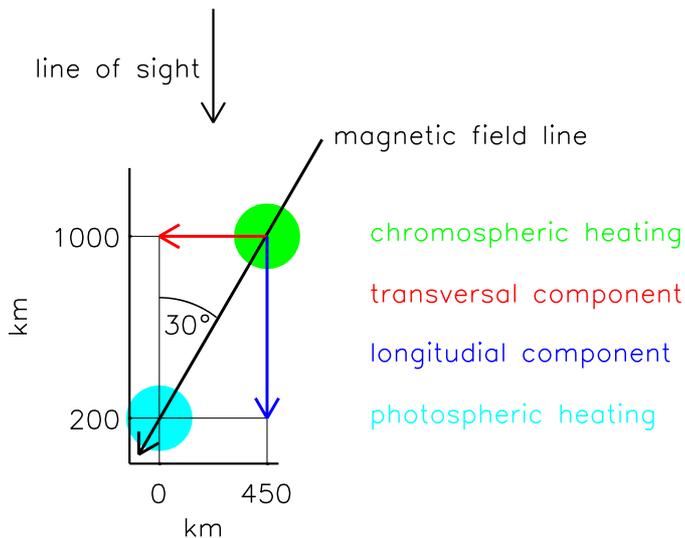}
 \caption{Simplified scheme of the beam energy deposition along a particular magnetic field line and its projection to the LOS frame. The flare ribbon axis is assumed to be perpendicular to the figure plane.} 
 \label{scheme}
\end{figure}

\begin{figure}[!t]
 \centering \includegraphics[width=\linewidth]{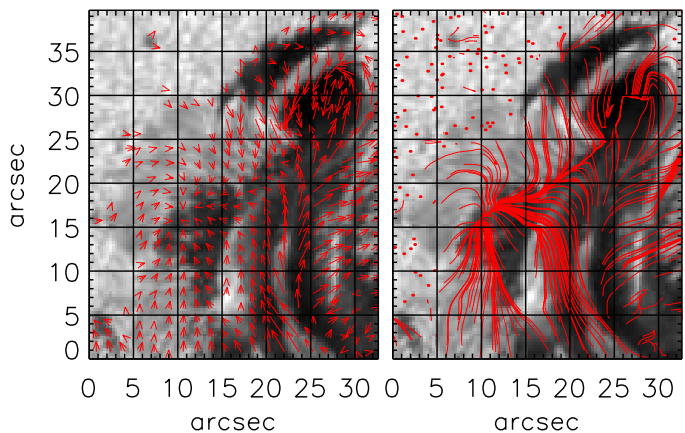}
 \caption{Intensity maps overplotted with the arrows indicating the orientation and strength of the transversal component of the magnetic field (left) and with the streamlines of this magnetic field component (right).}
 \label{streamlines}
\end{figure}

\begin{figure*}[!t]
 \centering \includegraphics[width=0.9\linewidth]{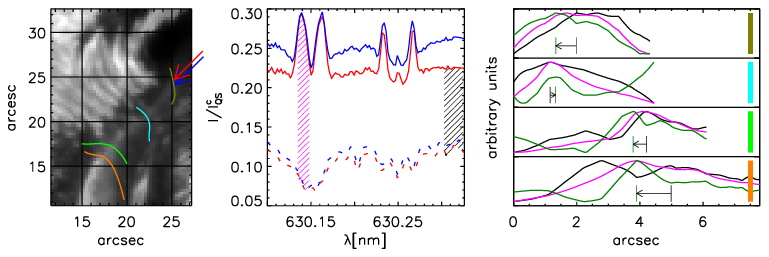}
 \caption{Left panel:  Continuum intensity map with the streamlines of the transversal component of the magnetic field indicated by orange, green, cyan, and olive lines. The blue and red arrows indicate the pixels displayed in the middle panel. Middle panel:  Stokes $I$ profiles observed during the flare (solid lines) and post-WLF phase (dashed lines). The shaded areas give the range of wavelengths  used to compute the average intensity difference between the Stokes $I$ profiles observed during the flare and post-WLF phase. Right panel: Continuum and line intensifications (black and pink lines) for pixels along the streamlines plotted in the left panel; the green lines correspond to the difference. For display purposes, the maxima and minima of these lines are normalised to the same range. The thick coloured vertical lines on the right  of the plots show the respective streamline colours. The thin vertical black and green lines show the local maxima of black and green lines, respectively. The arrows connecting them indicate the shift of these maxima. The zero on the $x$-axis corresponds to the upper-left ends of the streamlines in the left panel.} 
 \label{int_diff}
\end{figure*}

\subsection{Possible geometrical separation of continua-forming regions}

According to \citet{Heinzel:2017}, the chromospheric continuum is formed approximately 1000~km higher than the photospheric continuum. We assume that  the chromospheric continuum and the \ion{Fe}{i} line emission are both caused by the energy deposition from the electron beam that is propagating to the lower atmospheric layers along the magnetic field lines. A simplified scheme of this scenario is shown in Fig.~\ref{scheme}. Therefore, we might in principle detect an offset between the regions with maximum continuum-intensity enhancement (caused by the chromospheric continuum) and those with the profiles with the most significant emission characteristics (which are caused by a heated photosphere between $\log \tau =-1$ and $-2$ corresponding roughly to geometrical heights between 150~km and 300~km; see the left panel of Fig.~\ref{rf}).

We note that the heating of photospheric layers can be a combination of beam energy deposit and radiative backwarming by the chromospheric hydrogen Balmer continuum, the latter causing a temperature increase even in regions not directly affected by the beam itself \citep[for possible observational evidence, see e.g. ][]{1993ApJ...406..306N, 2006ApJ...641.1210X, 2007PASJ...59S.807I}. Therefore, the assumed offset can be biased by the photospheric continuum enhancement due to such backwarming.

In the left panel of Fig.~\ref{streamlines} we have indicated by arrows the transversal component of the magnetic field. The azimuth values of these arrows were used to construct the streamlines shown in the right panel of Fig.~\ref{streamlines}, i.e.  lines along which the electron beam would be propagating from the observers point of view. The ideal location to identify the spatial offset between the chromospheric and photospheric heating is where the streamlines are perpendicular to the ribbon orientation, i.e. around [19\arcsec, 18\arcsec]:  the magnetic field is inclined by 30$^\circ$ to the LOS and points from right to left towards the negative polarity region. The identified configuration of the magnetic field is in agreement with the scheme shown in Fig.~\ref{scheme}. The estimated height difference of 800~km would be shortened to approximately 450~km, i.e. around two pixels with our spatial sampling. This is a significantly smaller spatial displacement than the width of the observed flare ribbon. 

Examples of Stokes $I$ profiles that are in agreement with our expectations are shown in the middle panel in Fig.~\ref{int_diff}. The blue line depicts an emission profile with a continuum enhancement of $0.12 I^c_{QS}$ compared to the post-WLF phase. The red line shows an emission profile, where the emission characteristics are more pronounced as there is almost no decrease in intensity in the far wings of the \ion{Fe}{i} lines, and the continuum enhancement is $0.1 I^c_{QS}$ compared to the post-WLF phase. Presumably, the chromospheric heating is more pronounced in the pixel shown by the blue line and the photospheric heating is more pronounced in the pixel shown by a red line further along the same field line. The distance between these pixels is 0\farcs7, i.e. around 500~km, and corresponds to the expected displacement  estimated in Fig.~\ref{scheme}. We cannot expect to have completely separated characteristics of chromospheric and photospheric heating due to the width of the flare ribbon and the possible contribution caused by backwarming. 

We use the intensity difference in the continuum wavelengths between the flare and post-WLF phase as a proxy for chromospheric heating (black lines in the right panel of Fig.~\ref{int_diff}). As a proxy for the photospheric heating (green lines in the right panel of Fig.~\ref{int_diff}), we compute the intensity difference between the flare and post-WLF phase in the line wing (pink lines in the right panel of Fig.~\ref{int_diff}) and subtract the intensity difference at the continuum wavelengths as we expect a completely flat continuum contribution from the chromosphere. 

These proxy values are plotted in the right panel of Fig.~\ref{int_diff} along the streamlines of the transversal component of the magnetic field plotted in the left panel of Fig.~\ref{int_diff}. Local maxima of black lines depict the maximum of the proxy of the chromospheric heating and maxima of green lines show the regions with the maximum of the proxy of the photospheric heating. The arrows connecting these maxima show their spatial displacement. In three cases out of four, the arrow orientation is in agreement with our estimation of the spatial displacement of the photospheric and chromospheric heating. The spatial separation of the maxima is between 0\farcs5 and 1\arcsec. We note that the local maxima of the black lines for the orange and green streamlines located around 3\arcsec and 2\farcs5, respectively, are caused by the evolution of the light bridge intensity and are not related to the WLF.

Only in one case (cyan streamline) does the arrow point in a direction that is not in agreement with our interpretation of the projection effect. It should be  noted that although the results presented in Fig.~\ref{int_diff} are mostly in agreement with our expectations, they cannot be taken as confirmation of the projection effect. The used proxies do not represent the chromospheric and photospheric heating accurately. The magnetic field configuration is more complex and not retrieved precisely with our height-independent model atmosphere. The energy deposition along the field lines is dependent on the physical properties of the chromosphere that are unknown and can differ from pixel to pixel. In addition, the location of the flare-loop footpoints where beams of accelerated particles propagated is not known because imaging hard X-ray data are not available during the time of the Hinode flare scan. Thus, the nature of the atmospheric heating cannot be identified either.

\section{Conclusions}
\label{conclussion}

We use a unique dataset obtained by the Hinode satellite capturing the X9.3-class WLF ribbon with a spectropolarimetric raster scan. The observed Stokes profiles  allow us to interpret the temperature stratification of the solar photosphere during this WLF under the assumption of LTE conditions. Comparing these data with temperatures obtained in the post-WLF phase, we can also estimate the heating of the photosphere during the WLF.

In the studied WLF ribbon, the flare affected also the upper photosphere. In pixels with the most enhanced emissions, the photosphere is heated down to $\log \tau = -0.5$, i.e. approximately 80~km above the solar surface. This is in agreement with the shape of the Stokes $I$ profiles in these pixels as  the \ion{Fe}{i} lines show absorption features in the far wings, i.e. it signifies a temperature decrease around $\log \tau = 0$.

Nevertheless, we observe enhancements of the continuum intensity equivalent to $0.3 I^c_{QS}$ that can be ascribed to the WLF ribbon. This corresponds in certain pixels to an increase of $I^c$ by 200\% (see Figs.~\ref{profile_comparison} and~\ref{histo}). Following the analogy of the study by \citet{Heinzel:2017}, we assume that this continuum contribution is mostly due to the Paschen recombination continuum formed in the chromosphere. However, we are aware that there are studies locating the WLF sources at various layers in the photosphere \citep{2012ApJ...753L..26M} and in the chromosphere \citep{kruckeretal2015,Heinzel:2017}. 

If chromospheric contribution to the continuum intensity is not taken into account and the observed Stokes profiles are used to estimate the temperature stratification in the atmosphere by means of inversion codes, the resulting temperatures are overestimated by approximately 6\% for the lowest values and by 17\% for the highest values (left plot of Fig.~\ref{scatterplot}). Taking the chromospheric contribution into account is also necessary to estimate the actual heating of the upper photosphere in comparison to the post-WLF phase. In the case of the studied flare, we found the maximum heating at $\log \tau = -2$ around 1500~K, but most common are values between 300~K and 1000~K (right plot of Fig.~\ref{scatterplot}).

Our analysis also shows that the photosphere contributes to the increased continuum intensity. However, this contribution is small compared to the chromospheric continuum and the derived values are typically around $0.01 I^c_{QS}$ with maximum of $0.03 I^c_{QS}$ (see displacement of histograms in Fig.~\ref{histo}). We note that the values of the chromospheric continuum represent upper estimates because we do not account for backwarming and other possible mechanisms that can increase the continuum intensity and do not originate in the chromosphere. 

We find hints of a spatial displacement of the location of maximum heating of the chromosphere and upper photosphere. These displacements are in agreement with the projection effects expected from the magnetic field configuration. However, the found values of displacement are on a pixel-size scale and smaller than the width of the flare ribbon, and need to  be confirmed by other observations with better spatial resolution along with observations of spectral lines forming in the chromosphere.

We did not find any significant impact of the WLF on the LOS~velocities observed in the analysed region. The maximum $v_\mathrm{LOS}$ values are around 0.5~km~s$^{-1}$ (right panel of Fig.~\ref{velocity_maps}). However, we specifically focused on regions with symmetric Stokes profiles where using the  simple model of the atmosphere is justifiable. There are hints of enhanced LOS~velocities in the flare ribbon around [16\arcsec, 4\arcsec$-$12\arcsec];  the Stokes profiles are more asymmetric and the LOS~velocities might be influenced by both the WLF and by the sunspot light bridge, i.e. more careful analysis of $v_\mathrm{LOS}$ is needed to investigate these regions.

We are aware that the inversion code SIR works under the assumption of LTE and hydrostatic equilibrium and may not be well suited for the analysis of a flaring atmosphere, primarily because of the expected chromospheric contribution to the continuum intensity. Although the observed Stokes profiles are convincingly fitted under these assumptions, the results are biased (see explanation in Sect.~\ref{discussion}). A self-consistent non-LTE RHD model is needed to properly describe the physical conditions in a flaring atmosphere and to confirm the credibility of the presented results. This is beyond the  scope of the presented paper, and will be a subject of a future study.

\begin{acknowledgements}

We thank Hugh Hudson for the comments and discussions on the paper. The authors were supported by the Czech Science Foundation under grants 18-06319S (J.J. and M.\v{S}.) and 16-18495S (J.K.). J.J., J.K., and M.\v{S}. acknowledge the support from RVO:67985815. Hinode is a Japanese mission developed and launched by ISAS/JAXA, with NAOJ as domestic partner and NASA and STFC (UK) as international partners. It is operated by these agencies in cooperation with ESA and NSC (Norway). 

\end{acknowledgements}

\bibliographystyle{aa}
\bibliography{manuscript}

\end{document}